# Elemental topological Dirac semimetal: $\alpha$-Sn on InSb(111)

Cai-Zhi Xu[1,2,3], Yang-Hao Chan[4], Yige Chen[5], Peng Chen[1,2,3], Xiaoxiong Wang[1,2,6], Catherine Dejoie[7], Man-Hong Wong[1,2], Joseph Andrew Hlevyack[1,2], Hyejin Ryu[3], Hae-Young Kee[5], Nobumichi Tamura[3], Mei-Yin Chou[4,8,9], Zahid Hussain[3], Sung-Kwan Mo[3] * and Tai-Chang Chiang[1,2,9] *

[1]Department of Physics, University of Illinois at Urbana-Champaign, Urbana, Illinois 61801, USA

[2]Frederick Seitz Materials Research Laboratory, University of Illinois at Urbana-Champaign, Urbana, Illinois 61801, USA

[3]Advanced Light Source, Lawrence Berkeley National Laboratory, Berkeley, California 94720, USA

[4]Institute of Atomic and Molecular Sciences, Academia Sinica, Taipei 10617, Taiwan

[5]Department of Physics, University of Toronto, Toronto, Ontario M5S 1A7, Canada

[6]College of Science, Nanjing University of Science and Technology, Nanjing 210094, China

[7]Structure of Materials Group, ESRF-The European Synchrotron CS40220, 38043 Grenoble Cedex 9, France

[8]School of Physics, Georgia Institute of Technology, Atlanta, Georgia 30332, USA

[9]Department of Physics, National Taiwan University, Taipei 10617, Taiwan

*Correspondence and requests for materials should be addressed to S.-K.M. (email: SKMo@lbl.gov) or T.-C.C. (email: tcchiang@illinois.edu).*

**Abstract**

Three-dimensional (3D) topological Dirac semimetals (TDSs) are rare but important as a versatile platform for exploring exotic electronic properties and topological phase transitions. A quintessential feature of TDSs is 3D Dirac fermions associated with bulk electronic states near the Fermi level. Using angle-resolved photoemission spectroscopy (ARPES), we have observed such bulk Dirac cones in epitaxially-grown $\alpha$-Sn films on InSb(111), the first such TDS system realized in an elemental form. First-principles calculations confirm that epitaxial strain is key to the formation of the TDS phase. A phase diagram is established that connects the 3D TDS phase through a singular point of a zero-gap semimetal phase to a topological insulator (TI) phase. The nature of the Dirac cone crosses over from 3D to 2D as the film thickness is reduced.

Recent years have witnessed an ever-growing interest in materials that host massless Dirac fermions; examples include graphene [1], TIs [2] and Dirac semimetals [3–8]. Three-dimensional (3D) topological Dirac semimetals (TDSs) have recently attracted much attention as the first example of topological phases of gapless systems that are the 3D counterpart of graphene [3,5–8]. In such systems, the bulk valence band (VB) and conduction band (CB) touch at multiple discrete points in the bulk Brillouin zone to form Dirac points, near which the dispersion relations are linear in all three momentum directions [3,4]. The bulk Dirac cones are protected by crystalline symmetry and spin-orbit coupling [4–7,9]. Furthermore, these materials can support nontrivial surface states [5–7,10]. The unusual electronic structure leads to extraordinary physical properties, including giant linear magnetoresistance [11,12], ultrahigh carrier mobility [12,9], chiral anomaly [13] and novel quantum oscillations [14]. Theoretically, TDSs can be driven by symmetry breaking into TIs and Weyl semimetals [3,5,13]; thus, these materials are excellent parent materials for realizing other exotic electronic phases. They also provide an excellent test ground for studies of topological phase transitions. TDSs are rare, however, because the topological conditions on the electronic structure are generally hard to satisfy. Thus far, $Na_3Bi$ and $Cd_3As_2$ are the only two known cases [3,8–10,15].

$\alpha$-Sn is a well-known semimetal with a zero-gap protected by the cubic symmetry of its crystal lattice [16]. A prior experimental study reported that a band gap opened in $\alpha$-Sn when compressive strain was applied along (111) or (001) direction [17], which broke the cubic symmetry. Liang Fu *et al.* theoretically proposed that such an insulator phase was a strong 3D topological insulator [18]. Two recent experimental studies demonstrated the existence of topological surface states in the valence band regions of $\alpha$-Sn films grown on InSb(001), but no experimental evidence of the strain-induced gap [19,20].

Herein, we show a surprising discovery that $\alpha$-Sn can be made into a TDS under a suitable strain achieved experimentally through molecular beam epitaxial (MBE) growth of $\alpha$-Sn thin films on InSb(111). Using ARPES, we have directly observed a Dirac cone with linear band dispersion relations along all three momentum directions, substantiated by our first-principles calculations. As the film thickness is reduced the character of this Dirac cone crosses over from 3D to 2D. In addition, our first-principles calculations demonstrate that $\alpha$-Sn can be a topological insulator or a TDS depending on the sign of applied strain, solving the discrepancy between our discovery and the prior reports [17].

The substrates for $\alpha$-Sn thin film growth were InSb(111)-B wafers cleaned by repeated cycles of Ar$^+$ sputtering and annealing at 400 °C. Thin films of $\alpha$-Sn were grown *in situ* by evaporation of Sn at a rate of ~3 minutes per bilayer onto a substrate kept at room temperature. Doping of the surface with K was performed with a SAES Getter dispenser. The ARPES measurements were performed at beam line 10.0.1 of the Advanced Light Source. All data were taken with a Scienta R4000 electron analyzer with the sample maintained at 40 K, with an energy resolution of 16 meV and angular resolutions of 0.1°. First-principles calculations were performed using the ABINIT code [21] and HGH-type pseudopotentials [22]. Spin-orbit coupling was taken into account. The exchange-correlation functional was obtained with the modified Becke-Johnson Local Density Approximation (MBJLDA) method [23], which yields a correct conduction band edge of $\alpha$-Sn at the *L* point [24]. The cutoff of electron kinetic energy was 340 eV. A *k*-space grid of 12×12×8 was adopted based on the Monkhorst-Pack algorithm. The lattice constant of bulk $\alpha$-Sn was set as the experimental value $a = 6.4892$ Å [28].

While bulk $\alpha$-Sn, with the diamond lattice structure, is stable only below 13 °C, thin films of $\alpha$-Sn grown on InSb can be stable up to ~170 °C [30]. The lattices of $\alpha$-Sn and InSb are nearly matched, but a slight mismatch results in an in-plane compressive strain of 0.14% for the $\alpha$-Sn overlayer [19,31]. The crystal structure of (111)-oriented $\alpha$-Sn films is shown in Fig. 1(a). It consists of a stack of bilayers (BLs). Each BL is a stanene, which resembles graphene but is buckled with the two triangular sublattices at different heights [32]. Fig. 1(b) shows the hexagonal prismatic bulk Brillouin zone and the (111)-projected surface Brillouin zone. A reflection-high-energy-electron-diffraction (RHEED) pattern of a 30-BL $\alpha$-Sn(111) film grown on InSb(111) (Fig. 1(c)) demonstrates a good film quality. The pattern also reveals a 3×3 reconstruction, which is known to exist on the $\alpha$-Sn(111) face [30]. Fig. 1(d) shows RHEED intensity as a function of deposition time.

The periodic oscillatory behavior indicates a layer-by-layer growth mode and permits precise determination of the film thickness. Fig. 1(e) shows a photoemission spectrum of the 30-BL film; it is dominated by the Sn 4$d$ core level doublet. No signals from the substrate In or Sb are detected; thus the film is continuous with no pin holes or cracks. In contrast to prior experiments that introduced Te [19] or Bi [20] into Sn films to help films grow smoothly, no such impurity atoms were introduced in our thin film samples yet they were of high-quality. The strain in the 30-BL film was measured by X-ray diffraction using the $sin^2\psi$ technique [24]. The results, shown in Fig. 1(f), yield an in-plane compressive strain $\varepsilon_\parallel$ = (-0.14 ± 0.03)%, which indicate that the $\alpha$-Sn film is fully epitaxially constrained to the InSb(111) in-plane lattice constant. The perpendicular strain from the experiment is $\varepsilon_\perp$ = (+0.006 ± 0.015)%, which is essentially zero within experimental error.

Fig. 2(a) shows the calculated band structure of unstrained bulk $\alpha$-Sn. Its CB and VB touch at the zone center, resulting in a single-point Fermi surface. The CB and VB nearby have quadratic dispersions; thus, the system is not a Dirac semimetal but an ordinary zero-gap semimetal [4]. This particular band topology is protected by the cubic symmetry of the system, but can be readily modified by uniaxial strain [17–19]. Note that the order of the $\Gamma_8^+$ CB and the $\Gamma_7^-$ VB in $\alpha$-Sn is inverted; modifications to the band topology can lead to non-trivial topological phases [18,19]. The theoretical band structure of bulk $\alpha$-Sn under the experimentally observed strain ($\varepsilon_\parallel = -0.14\%$) is shown in Fig. 2(b) for an overview and in Figs. 2(c)-(d) for detailed views. The strain causes the CB and VB to move closer together, resulting in a small (negative) gap of $\Delta$ ~20 meV at the Γ point. As shown in Fig. 2(g), the single-point Fermi surface in the unstrained case now splits into two points at (0, 0, ±$k_D$), with $k_D$ = 0.017(2π/c). The system remains a semimetal. The dispersion relations near the two contact points are linear in all three momentum directions; these features are indicative of a TDS. Given that the inverted band order in the unstrained system is preserved under strain, the 2D $Z_2$ index on the $k_z = 0$ plane is +1 [18]. Following the usual classification criteria [7], strained $\alpha$-Sn is indeed a TDS. The two bulk Dirac points, BDP1 and BDP2 as indicated in Fig. 2(g) are protected by the three-fold rotational symmetry in the strained structure, although the value of $k_D$ depends on the strain [24].

This is the first report of a TDS phase in α-Sn. Prior studies have reported instead a TI phase under strain [17,19]. The differences can be attributed to either a different direction – (001) [19] or a different sign of the strain [17]. Fig. 2(e) shows a calculated bulk band structure of α-Sn with a positive strain of $\varepsilon_\parallel = 0.7\%$. An absolute gap of ~50 meV is obtained, and the system is a TI because of the band inversion. At an even larger $\varepsilon_\parallel$ of 1.5%, the system becomes an ordinary semimetal (OS) because of the band overlap, as shown in Fig. 2(f). The phase diagram in Fig. 2(h) summarizes the systematics; the system is a TDS for a negative strain, becomes a zero-gap semimetal at zero strain, transforms into a TI at positive strain and finally becomes an OS at a sufficiently large positive strain.

ARPES was employed to examine our negatively strained films. Fig. 3(a) shows a Fermi surface map of a 6-BL film. It is a single point at the zone center, which corresponds to the projection of the two bulk Dirac points of strained α-Sn. The in-plane band dispersion of the same film along $\bar{K}$-$\bar{\Gamma}$-$\bar{K}$ determined by ARPES mapping is shown in Fig. 3(b). Right below the Fermi level $E_F$ is a band that disperses linearly (red dashed lines) away from the single-point Fermi surface. The green lines indicate another valence band feature with linear band dispersions. For comparison, Fig. 3(c) shows calculated ARPES spectral functions deduced from the surface-projected density of states of a semi-infinite α-Sn slab. The major features agree well with experiment. The red lines in Fig. 3(b) correspond to the top valence band. The green lines are associated with a Dirac point at about 0.3 eV binding energy; the relevant states have a strong surface character and originate from a topological surface band, but they become surface resonances because of the presence of degenerate bulk states. This surface resonance state is very similar to that on the (001) surface of $Cd_3As_2$ [6].

In order to view the dispersion relations of the CB, which normally sits above the Fermi level, we use potassium (K) for electron doping of the surface [3,8]. Figs. 3(d)-(e) show a comparison of the electronic structure of a 6-BL $\alpha$-Sn film before and after surface deposition of K, respectively. The net effect is a shift of the Fermi level upward by 0.21 eV, thus revealing part of the CB as a V-shaped band. The dispersion relations are shown again, with an enlarged scale, in Fig. 3(f) for comparison with the computed projection of the bulk bands onto the $k_z = 0$ plane as presented in Fig. 3(g). Fig. 3(h) displays measured constant-energy contours at various binding energies. The contour is a circle at the high energy end, shrinks to a point at the Dirac point, and opens up at lower energies; this latter contour is strongly modulated in intensity and appears more like two spots. The modulation can be attributed to matrix element effects associated with the s-polarization geometry in our experiments [33]. The weak CB feature in Fig. 3(f) can be enhanced by taking the second derivative of the data; the results are shown as an intensity map in Fig. 3(i) and as a set of momentum distribution curves in Fig. 3(j). Curve fitting of the results yields peak positions as shown in Figs. 3(j)-(k). It clearly shows a V-shaped CB with linear dispersions. The Fermi velocity along $\overline{\Gamma K}$ is 7.09 eV·Å or $1.07 \times 10^6$ m·s$^{-1}$ for the CB and 2.16 eV·Å or $3.26 \times 10^5$ m·s$^{-1}$ for the VB. These values are comparable to that of a high-mobility TDS $Cd_3As_2$ [8,9].

ARPES measurements have also been carried out with different photon energies to map out the band dispersion relations along $k_z$. Some of the key results obtained for 6-, 10- and 30-BL $\alpha$-Sn films doped with K are shown in Fig. 4(a). Corresponding computed band dispersion relations at various $k_z$ are shown in Fig. 4(b). The CB shifts to partly below the Fermi level by K doping. However, this energy shift is reduced as the film thickness increases because the electrons from K doping are diluted throughout the film thickness. As a result, the visible part of the CB is very much reduced for the 30-BL film. Evidently, the measured band dispersion relations for the 6- and 10-BL films do not change with photon energy, while those of the 30-BL film show significant variations. This different behavior is highlighted in Fig. 4(c), which plots the top of the VB at the zone center as a function of both photon energy and film thickness. Thus, the 30-BL film is characterized by 3D band dispersions with substantial variations along $k_z$, while thinner films are characterized by 2D band dispersions.

Shown in Fig. 4(b) are computed in-plane band dispersion relations along $k_x$ for bulk $\alpha$-Sn at various values of $k_z$ chosen to correspond closely to the 30-BL data in Fig. 4(a). These $k_z$ values are also consistent with the photon energies chosen for the experiment based on a free-electron final band dispersion [34]. When $k_z = k_D$ (42 eV photon energy), the calculated band structure shows a Dirac point with Λ-shaped linear dispersion relations for the VB, in agreement with experiment [24]. As $k_z$ moves away, a gap develops between the VB and CB, again in agreement with experiment. Fig. 4(d) shows the experimental $k_z$ dispersion relations of bulk $\alpha$-Sn. Around $k_z = k_D$ (42 eV photon energy), a linear band dispersion can be seen. Note that the resolution of band mapping along $k_z$ is inherently limited by a finite momentum resolution due to a finite mean free path of the photoelectrons [34]. Thus the two bulk Dirac points predicted by first-principles calculation, being very close along $k_z$ axis, are not resolved in the experimental data. We emphasize that the observed Dirac cone with linear in-plane and out-of-plane dispersions near Fermi level is only consistent with the calculated band structure of $\alpha$-Sn with negative in-plane strain, which has been well confirmed by our X-ray diffraction measurements.

The 2D character of the band dispersions for the 6- and 10-BL films can be understood as a result of quantum confinement [35]. The bulk bands are discretized into quantum well states or subbands characterized by subband indices related to specific $k_z$ values. The Dirac cones seen for the 6- and 10-BL films are 2D Dirac cones, similar to those observed in graphene [1] or the topological surface states in 3D topological insulators [2]. However, the Dirac cones in the Sn films are derived from 3D electronic states. The differences in the results between the 30-BL film and the thinner films indicate a crossover from a 3D Dirac semimetal to a 2D Dirac semimetal. Unlike graphene of which the 2D Dirac cone will open a band gap in the presence of significant spin-orbit coupling [36], this 2D Dirac semimetal is robust against such coupling effect because of the intrinsic strong spin-orbit interaction already present in Sn. Such feature gives tunability to the 2D Dirac semimetal phase in $\alpha$-Sn films that isn't attainable in graphene, such as the transition into a topological insulator under strain [37].

Our results establish the first known case of a 3D TDS based on a simple elemental material. Moreover, our theoretical results show that strain engineering can be an effective way to create novel phases from ordinary materials. Note that C (diamond), Si, Ge and $\alpha$-Sn are isoelectronic and share the same crystal structure, but only $\alpha$-Sn can be driven into a TDS phase by strain. This unique behavior is because the large spin-orbit coupling in Sn leads to an inverted band ordering in the parent phase. The symmetry-required contact point between the CB and VB can transform into a positive or negative gap by the application of strain. Such topological transformations then lead to either a TI or a TDS, under a positive or negative strain, respectively. Prior studies explored only the positive strain side of the phase diagram. Thinner $\alpha$-Sn films are 2D Dirac semimetals; the Dirac states are very similar to those seen in graphene, suggesting many possible applications based on existing graphene research. Our work indicates that $\alpha$-Sn is a promising material for device applications based on its rich topological phase diagram.

This work is supported by the U.S. Department of Energy (DOE), Office of Science (OS), Office of Basic Energy Sciences, Division of Materials Science and Engineering, under Grant No. DE-FG02-07ER46383 (T.C.C.) and the National Science Foundation under Grant No. EFMA-1542747 (M. Y. C.). BL10.0.1 (ARPES) & BL12.3.2 (Diffraction) at the Advanced Light Source is supported by the Office of Basic Energy Sciences of the U.S. DOE under Contract No. DE-AC02-05CH11231. C.-Z. X. is partially supported by the ALS Doctoral Fellowship in Residence. Y.-H. Chan is supported by a Thematic Project at Academia Sinica. Y. C. and H.-Y. K. are supported by Natural Sciences and Engineering Research Council of Canada. X. W. is supported by the National Science Foundation of China under Grant No. 11204133. H. R. acknowledges support from Max Planck Korea/POSTECH Research Initiative of the NRF under Project No. 2016K1A4A4A01922028.

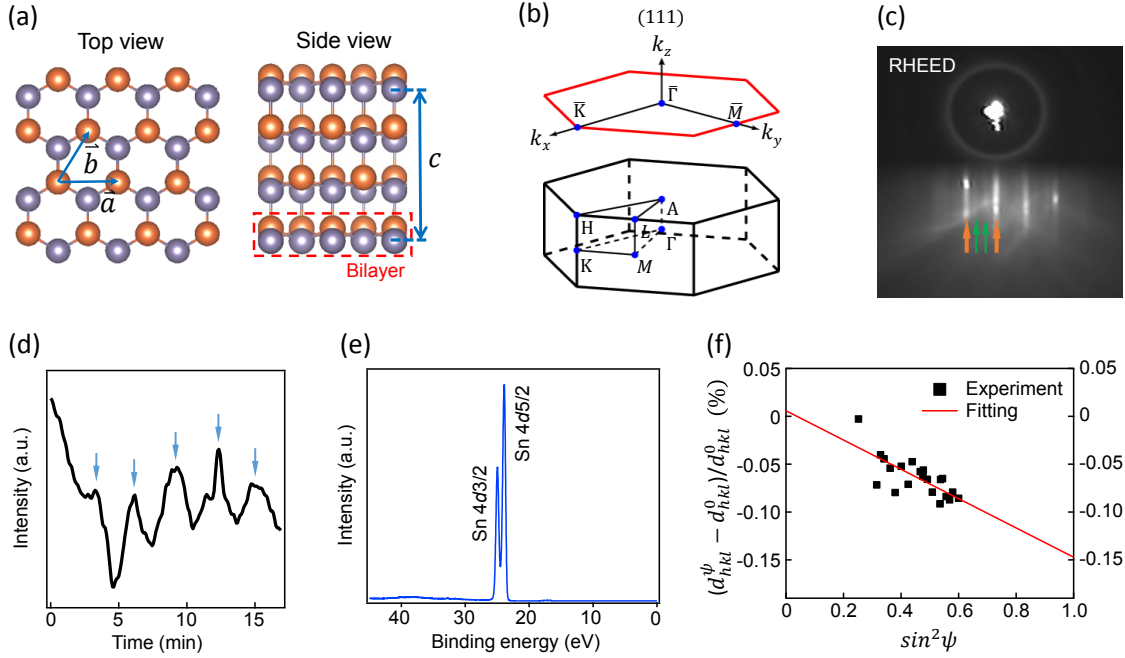

FIG. 1 (Color online). Crystal structure, RHEED pattern and strain analysis of α-Sn films on InSb(111). (a) Top and side views of the crystal structures of α-Sn films. The red dashed rectangle in side view shows the bilayer (BL) unit of α-Sn. (b) Bulk Brillouin zone and (111)-projected surface Brillouin zone of α-Sn. (c) A RHEED pattern of a 30-BL α-Sn film. The orange (green) arrows mark main (fractional) streaks. (d) RHEED intensity as a function of growth time. The blue arrows mark when each BL is formed. (e) A core-level photoemission spectrum for a 30-BL α-Sn film shows the characteristic Sn 4$d$ doublet. (f) Strain analysis of a 30-BL α-Sn film based on $sin^2\psi$ technique. $d'_{hkl}$ ($d^0_{hkl}$) is the interplanar spacing for the (hkl) plane of the measured α-Sn film (unstrained α-Sn). $\psi$ is the angle of each (hkl) plane with respect to the film surface plane.

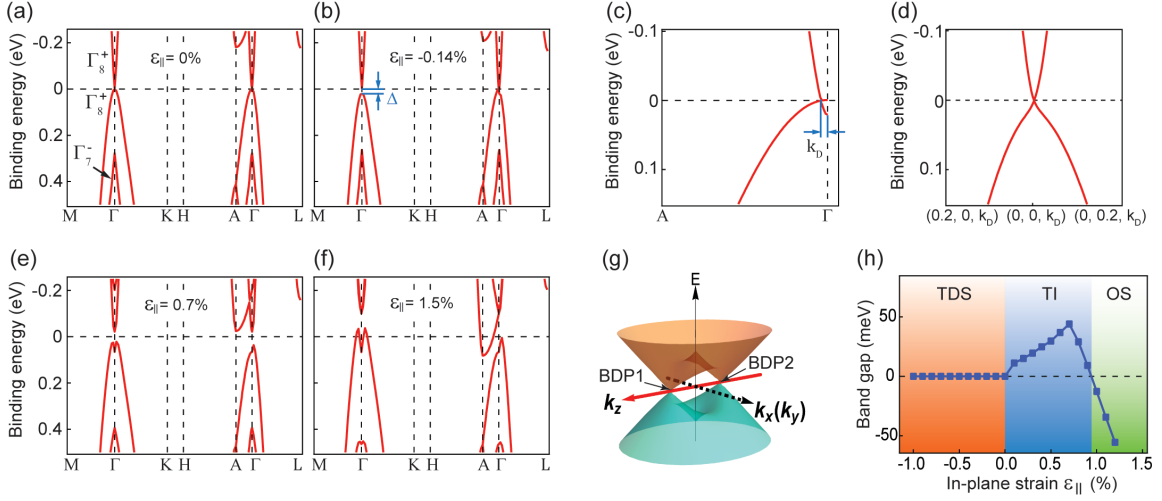

FIG. 2 (color online). Calculated electronic structure of $\alpha$-Sn and its dependence on strain. (a) Bulk band structure of $\alpha$-Sn without strain. (b) Bulk band structure of $\alpha$-Sn with the same strain ($\varepsilon_\parallel = -0.14\%$) as that observed experimentally. (c) Close-up view of (b) along the $\Gamma$A direction ($k_z$ direction). The wave number of the band crossing point is $k_D = 0.017(2\pi/c)$. (d) Bulk band structure of strained $\alpha$-Sn along the $k_x$ and $k_y$ directions near the bulk Dirac point. (e)-(f) Same as (b), but with in-plane strain $\varepsilon_\parallel$ different in signs and magnitude as indicated. (g) schematic illustration of the 3D TDS electronic structure of strained $\alpha$-Sn. BDP1 and BDP2 marks the positions of two bulk Dirac points. (h) Phase diagram and band gap of $\alpha$-Sn as a function of in-plane strain $\varepsilon_\parallel$. The band gap is defined as $E_c - E_v$, where $E_c$ ($E_v$) is the band edge of conduction (valence) band.

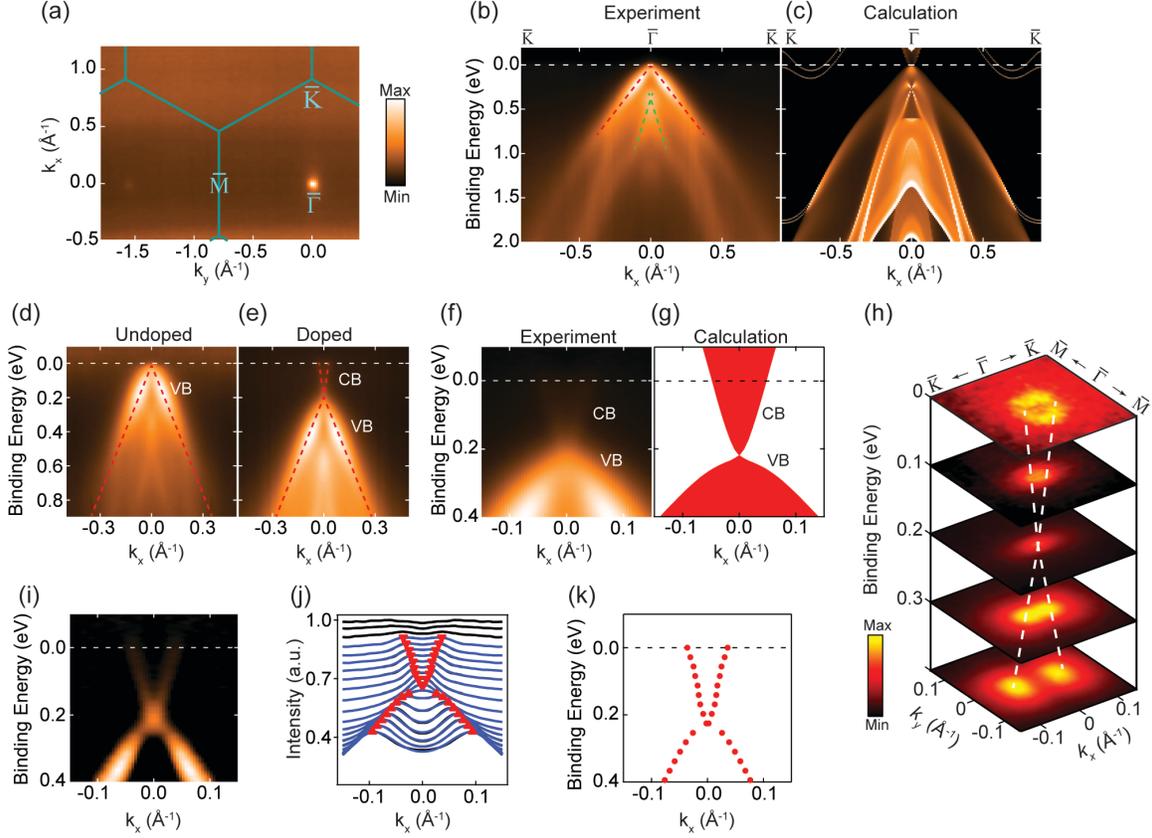

FIG. 3 (color online). Electronic structure of a 6-BL $\alpha$-Sn film. (a) Experimental Fermi surface. (b) Experimental band dispersion along $\bar{K}-\bar{\Gamma}-\bar{K}$. (c) Calculated ARPES spectral function of a semi-infinite $\alpha$-Sn slab in the same scale as that in (b). (d)-(e), Band dispersion before and after *in situ* electron doping using a potassium metal dispenser, respectively. (f) Close-up view of band dispersion in e around the zone center near the Fermi level. (g) Calculated bulk bands projected onto (111) for comparison with (f). (h) Stacking plots of constant energy contours at different binding energies to show a Dirac cone. (i) Second-derivative plot of experimental data shown in (f). (j) Momentum distribution curves obtained from experimental data in (f) (black curves) and peak fitting results (blue curves); these curves are essentially indistinguishable. Red triangles mark the peak positions for each blue curve. (k) Peak positions from fitting results show a Dirac cone.

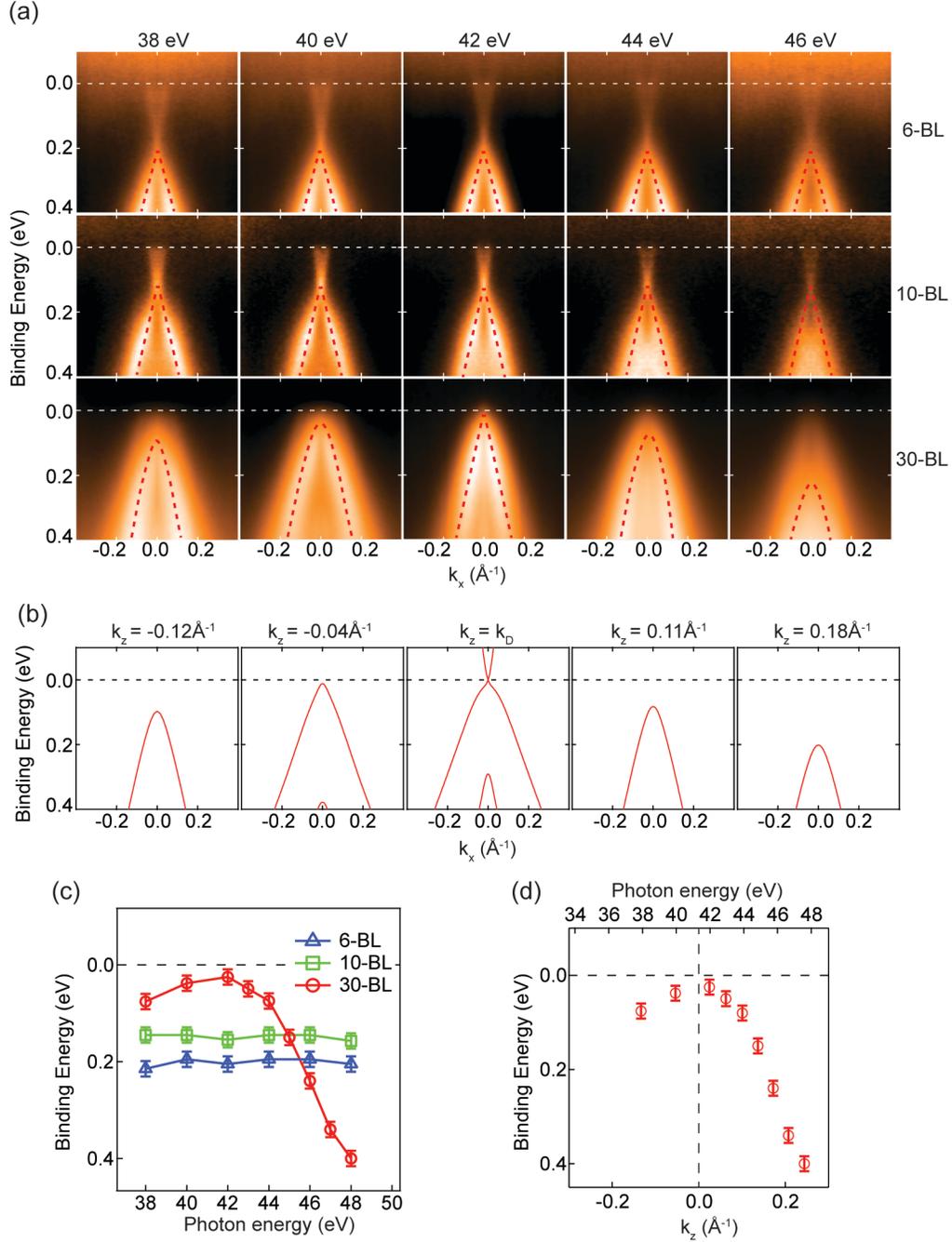

FIG. 4 (Color online). Evolution of band dispersion of $\alpha$-Sn films with film thickness and incident photon energy. (a) ARPES maps from 6-, 10- and 30-BL $\alpha$-Sn films under selected incident photon energies. (b) Calculated band dispersions of bulk $\alpha$-Sn along the $k_x$ axis at various out-of-plane wave vector $k_z$. (c) Evolution of the energy positon of the valence band top with incident photon energy for 6-, 10- and 30-BL $\alpha$-Sn films. (d) Band dispersions of a 30-BL $\alpha$-Sn film along the out-of-plane direction ($k_z$) from experiment.